\documentclass[aps,prb,a4paper,twocolumn,nobibnotes]{revtex4-1}
\usepackage{helvet}
\usepackage{graphicx}
\usepackage{xspace}
\usepackage{amsmath,amssymb,bm}
\usepackage[usenames,dvipsnames]{xcolor}

\usepackage{MnSymbol}

\newif\ifdrafttext
\drafttextfalse 
\ifdrafttext \usepackage[colorlinks,urlcolor=black,citecolor=black,linkcolor=black]{hyperref} \else   \usepackage[hidelinks]{hyperref} \fi

\makeatletter
\@addtoreset{figure}{myfigure}
\def\@caption@fignum@sep{~$\vert$~}%
\def\fnum@figure{\textbf{\figurename~\thefigure}}
\makeatother

\makeatletter
\@addtoreset{table}{mytable}
\def\@caption@tablenum@sep{~$\vert$~}%
\def\tnum@table{\textbf{\tablename~\thetable}}
\makeatother

\newcommand{\nuke}[1]{}

\newcommand{\note}[1]{{\color{red}{#1}}}
\newcommand{\abstracta}[1]{\textcolor{red}{#1}}
\newcommand{\abstractb}[1]{\textcolor{blue}{#1}}
\newcommand{\abstractc}[1]{\textcolor{ForestGreen}{#1}}
\newcommand{\abstractd}[1]{\textcolor{Fuchsia}{#1}}
\newcommand{\abstracte}[1]{\textcolor{Magenta}{#1}}
\newcommand{\abstractf}[1]{\textcolor{SkyBlue}{#1}}

\newcommand{\titleofpaper}{Observation of quantum-limited heat conduction over macroscopic distances}
\newcommand{\QCDaffiliation}{QCD Labs, COMP Centre of Excellence, Department of Applied Physics, Aalto University, P.O. Box 13500, FI--00076 Aalto, Finland}

\ifdrafttext
\else
    \renewcommand{\abstracta}[1]{#1}
    \renewcommand{\abstractb}[1]{#1}
    \renewcommand{\abstractc}[1]{#1}
    \renewcommand{\abstractd}[1]{#1}
    \renewcommand{\abstracte}[1]{#1}
    \renewcommand{\abstractf}[1]{#1}
\fi

\begin{document}

\title{\titleofpaper}

\author{Matti Partanen}

\author{Kuan Yen Tan}

\author{Joonas Govenius}

\author{Russell E.  Lake}

\author{Miika K. M\"akel\"a}

\author{Tuomo Tanttu}

%

\author{Mikko M\"ott\"onen}
\affiliation{\QCDaffiliation}

\date{\today}

\begin{abstract}

\abstracta{The emerging quantum technological apparatuses\cite{Wolf_kirja, Dowling2003}, such as the quantum computer\cite{Ladd,Clarke,Morton2011}, call for extreme performance in thermal engineering at the nanoscale\cite{Giazotto}.
}
\abstractb{Importantly, quantum mechanics sets a fundamental upper limit for the flow of information and heat, which is quantified by the quantum of thermal conductance\cite{Pendry, Rego}. 
The physics of this kind of quantum-limited heat conduction has been experimentally studied for lattice vibrations, or phonons\cite{Schwab}, for electromagnetic interactions\cite{Meschke_Nature}, and for electrons\cite{Jezouin}.
}
\abstractc{However, the short distance between the heat-exchanging bodies in the previous experiments hinders the applicability of these systems in quantum technology.
}
\abstractd{Here, we present experimental observations of quantum-limited heat conduction over macroscopic distances extending to a metre.
}
\abstracte{We achieved this striking improvement of four orders of magnitude in the distance by utilizing microwave photons travelling in superconducting transmission lines. 
Thus it seems that quantum-limited heat conduction has no fundamental restriction in its distance.
}
\abstractf{This work lays the foundation for the integration of normal-metal components into superconducting transmission lines, and hence provides an important tool for circuit quantum electrodynamics~\cite{Blais,Wallraff_Nature_2004, Astafiev2010}, which is the basis of the emerging superconducting quantum computer~\cite{Kelly2015}. 
In particular, our results demonstrate that cooling of nanoelectronic devices can be carried out remotely with the help of a far-away engineered heat sink.
}
{
In addition, quantum-limited heat conduction plays an important role in the contemporary studies of thermodynamics such as fluctuation relations and Maxwell's demon\cite{Pekola_2015, Golubev_2015}.
 Here, the long distance provided by our results may, for example, lead to an ultimate efficiency of mesoscopic heat engines with promising practical applications\cite{Sothmann}.
}

\end{abstract}


\maketitle



%
%
%
%
%
%
%
%

\newpage


The quantum of thermal conductance, $G_\textrm{Q}$, provides the fundamental upper limit for  heat conduction through a single channel \cite{Pendry}.
This limit applies to fermions and bosons as the carriers of heat as well as to so-called anyons obeying even more general statistics \cite{Rego}.
Although a few observations of quantum-limited heat conduction have been reported,  the studied distances were shorter than 100~$\mu$m in all previous experiments:
phononic heat conduction   through four parallel submicrometre dielectric wires  each supporting four vibrational modes \cite{Schwab},
electromagnetic heat conduction  in a superconducting loop over a 50-$\mu$m distance \cite{Meschke_Nature, Timofeev_1}, 
and  electronic heat conduction   through an extremely short quantum point contact engineered in a two-dimensional electron gas~\cite{Jezouin}.
Photons,  unlike  many other  carriers of heat, can travel macroscopic distances without significant scattering, for example, in optical fibres or superconducting waveguides. 
Thus, photons seem ideal for long-distance thermal engineering and provide attractive opportunities for various quantum thermodynamics experiments~\cite{Pekola_2015}.
To our knowledge, however, itinerant photons have not been previously employed in experimental studies of the quantum of thermal conductance.

\begin{figure}[t!]
  \centering
  \def\svgwidth{89mm}
  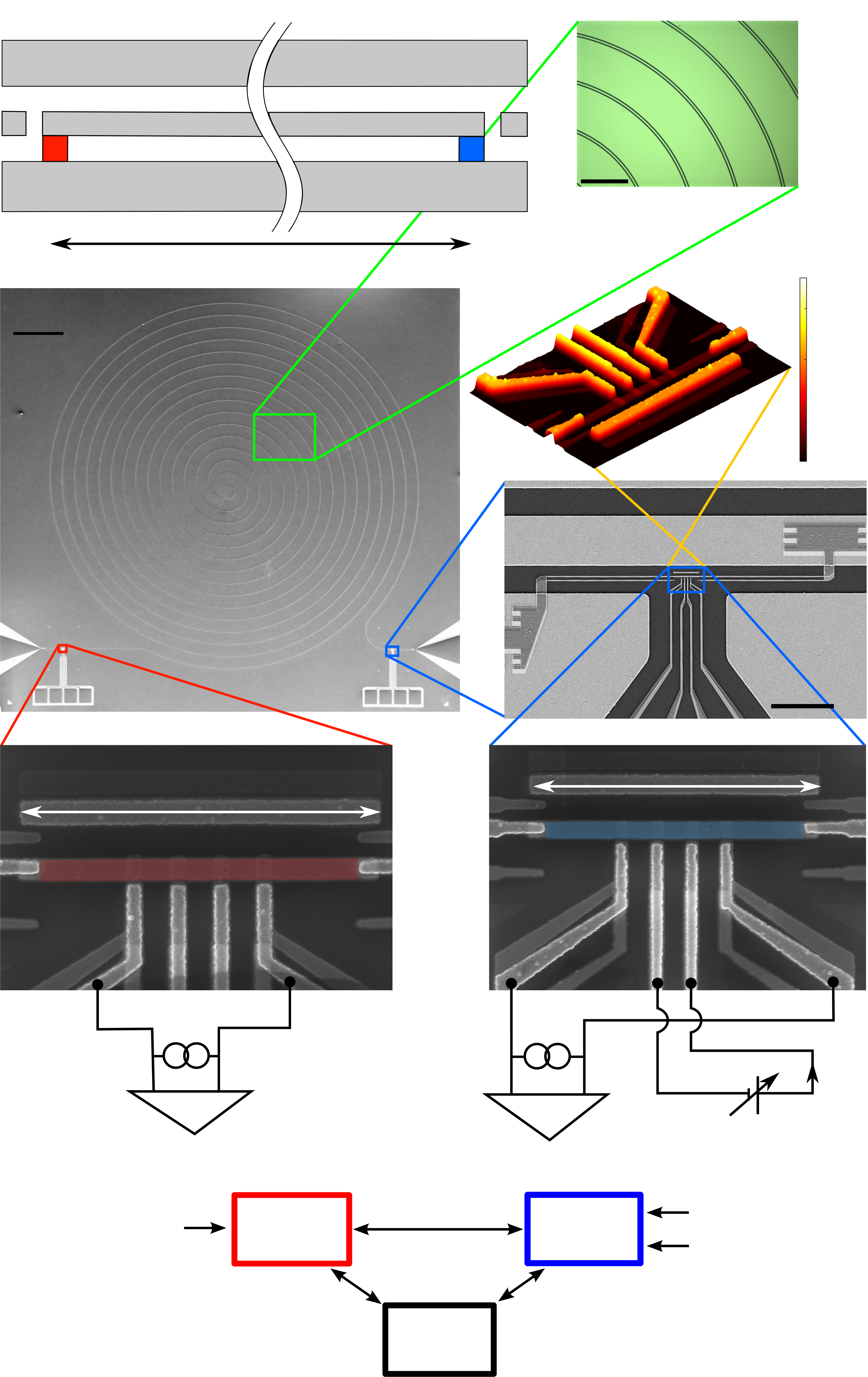
  \caption{ 
  \textbf{Sample structure and measurement scheme.}
   \textbf{a}, Schematic illustration of a coplanar  transmission line terminated at different ends by resistances $R_\textrm{A}$ and $R_\textrm{B}$ at temperatures $T_\textrm{A}$ and $T_\textrm{B}$, respectively.
   \textbf{b}, Scanning electrom microscope (SEM) image of a fabricated transmission line  with a double-spiral structure.
   \textbf{c, g},  False-colour SEM images of the normal-metal islands together with a simplified measurement scheme.
   \textbf{d}, Optical micrograph of the  waveguide.
   \textbf{e}, Atomic force microscope image of Island~B highlighting the thicknesses of the nanostructures.
   \textbf{f}, SEM image showing  how the normal-metal island is connected to the ground plane and to the centre conductor.
   \textbf{h}, Thermal model indicating the thermal conductance between the Islands A and B,  $G_\textrm{AB}$, and those from the islands to the phonon bath at the temperature $T_0$,  $G_\textrm{A0}$ and $G_\textrm{B0}$. 
   Constant powers $P_\textrm{const,A/B}$ and control power $P_\textrm{NIS}$ are also indicated by arrows.
  Micrographs  (\textbf{c, e, f, g}) are from Sample A1, and  (\textbf{b, d})  are from a similar sample.
  }
  \label{fig:sample_structure}
\end{figure}
\begin{figure}[t!]
  \centering
  \def\svgwidth{89mm}
  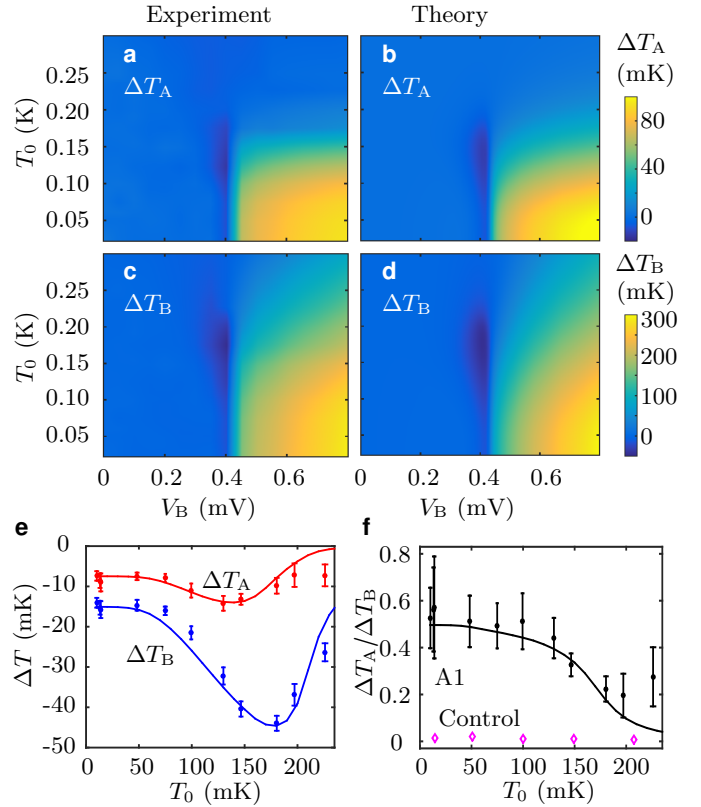
  \caption{ \textbf{ Photonic cooling at macroscopic distances for Sample A1. }
  \textbf{a, b}, Measured and theoretically predicted electron temperature changes with respect to the zero-bias case ($V_\textrm{B}=0$) for Island A as functions of the voltage $V_\textrm{B}$ and bath temperature $T_0$.
  \textbf{c, d}, As in panels (\textbf{a, b}), but for temperature changes of Island B. 
  \textbf{e},   Measured (markers) and simulated (lines) temperature changes at  the maximum cooling point as functions of the bath temperature. 
    The errorbars   indicate the standard deviation of the measured temperatures.
  \textbf{f}, The ratio of the temperature changes in (\textbf{e}). In addition, measurement results from the control sample are shown for comparison.  
  }
  \label{fig:full_simulation}
\end{figure}

In this Letter, we  experimentally study heat conduction through a single channel formed by  photons travelling in a long superconducting waveguide in a single transverse mode.
This scheme supports a  photonic thermal conductance very close to  $G_\textrm{Q}$. 
Since this heat transport does not directly depend on the  temperature of the substrate phonons, it provides an efficient method for remote temperature control. 
The  superconducting waveguide is terminated at both ends by resistors composed of mesoscopic normal-metal islands (Islands A and B  in Fig.~\ref{fig:sample_structure}). 
We measure the temperatures of the Islands A and B, and vary the temperature of Island~B.
A characteristic signal of photonic heat transport in our experiments is the increasing response  of the temperature of  Island~A to the controlled temperature changes of Island~B with decreasing phonon bath temperature (Figs.~\ref{fig:full_simulation} and \ref{fig:T2_simulation}).
The measured temperatures well agree with the thermal model, implying that the heat conduction essentially reaches the quantum limit.
Furthermore, the most important features of the measurement results can be quantitatively predicted from a theoretical model involving no free parameters (Fig.~\ref{fig:T2_simulation}).
These observations constitute firm evidence of quantum-limited heat conduction over macroscopic distances.

\begin{figure}[t!]
  \centering
  \def\svgwidth{89mm}
  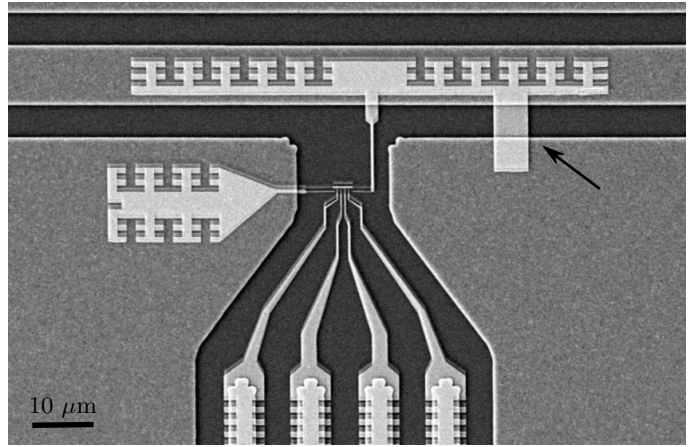
  \caption{{\bf Control sample.}  
  SEM image of the control sample. 
  The center conductor of the coplanar waveguide is shunted to the ground plane at the location indicated by the black arrow. 
  }
  \label{fig:Control_sample_SEM}
\end{figure}

Figure~\ref{fig:sample_structure} shows the structure of the sample   used in the experiments together with the measurement scheme.
We study several samples  with  different parameters as presented in Table \ref{tab:samples}. 
The length of the coplanar waveguide is either 20 cm or 1 m, and it has a double-spiral structure on a silicon chip with the size of   $1\times1$ cm$^2$ or $2\times2$ cm$^2$, respectively.
The electron temperatures of the  normal-metal islands are measured and controlled  with the help of normal-metal--insulator--superconductor (NIS)  junctions (Methods).
There are four nominally identical NIS junctions at each island.
In all samples, the normal-metal islands terminating the waveguide have two galvanic contacts to superconducting lines: one to the centre conductor of the waveguide and the other to the ground plane.
The waveguide has a centre conductor  width of 10~$\mu$m, and a separation between the ground plane and the centre conductor of 5~$\mu$m. 
In the control sample, the centre conductor is shunted to ground  to extinguish photonic heat conduction (Fig.~\ref{fig:Control_sample_SEM}).

\begin{table}[t!]
 \begin{center}
 \caption{ \textbf{Main parameters of the measured samples.} 
 Columns show the waveguide lengths, normal-metal resistances $R_i$, $i\in\{$A, B$\}$, normal-metal materials, and normal-metal volumes (length$\times$width$\times$thickness).
 Furthermore, $G_\Gamma/G_\textrm{Q}$ provides the estimated ratio of the realised photonic thermal conductance and the quantum of thermal conductance at temperatures of approximately 150 mK.}
 \begin{tabular}{ c | c |c | c | c | c }
   \hline
   Sample & Length (m)& $R_i$ ($\Omega$) & Material & Volume (nm$^3$) & $G_\Gamma/G_\textrm{Q}$ \\
   \hline
   \hline
   A1  	& 0.2	& 65 	& Cu 	& $5000 \times 300 \times 20 $ 	& 98 \%	\\
   A2  	& 1.0	& 75 	& AuPd	& $3200 \times 300 \times 40 $ 	& 94 \% \\
   A3  	& 0.2	& 150	& AuPd	& $3200 \times 300 \times 20 $ 	& 60 \% \\
   Control & 0.2& 100	& AuPd	& $3200 \times 300 \times 20 $ 	& 0 \% \\
\hline
\end{tabular}
 \label{tab:samples}
 \end{center}
\end{table}

\begin{figure}[t!]
  \centering
  \def\svgwidth{89mm}
  {\fontsize{10pt}{10pt}\selectfont
  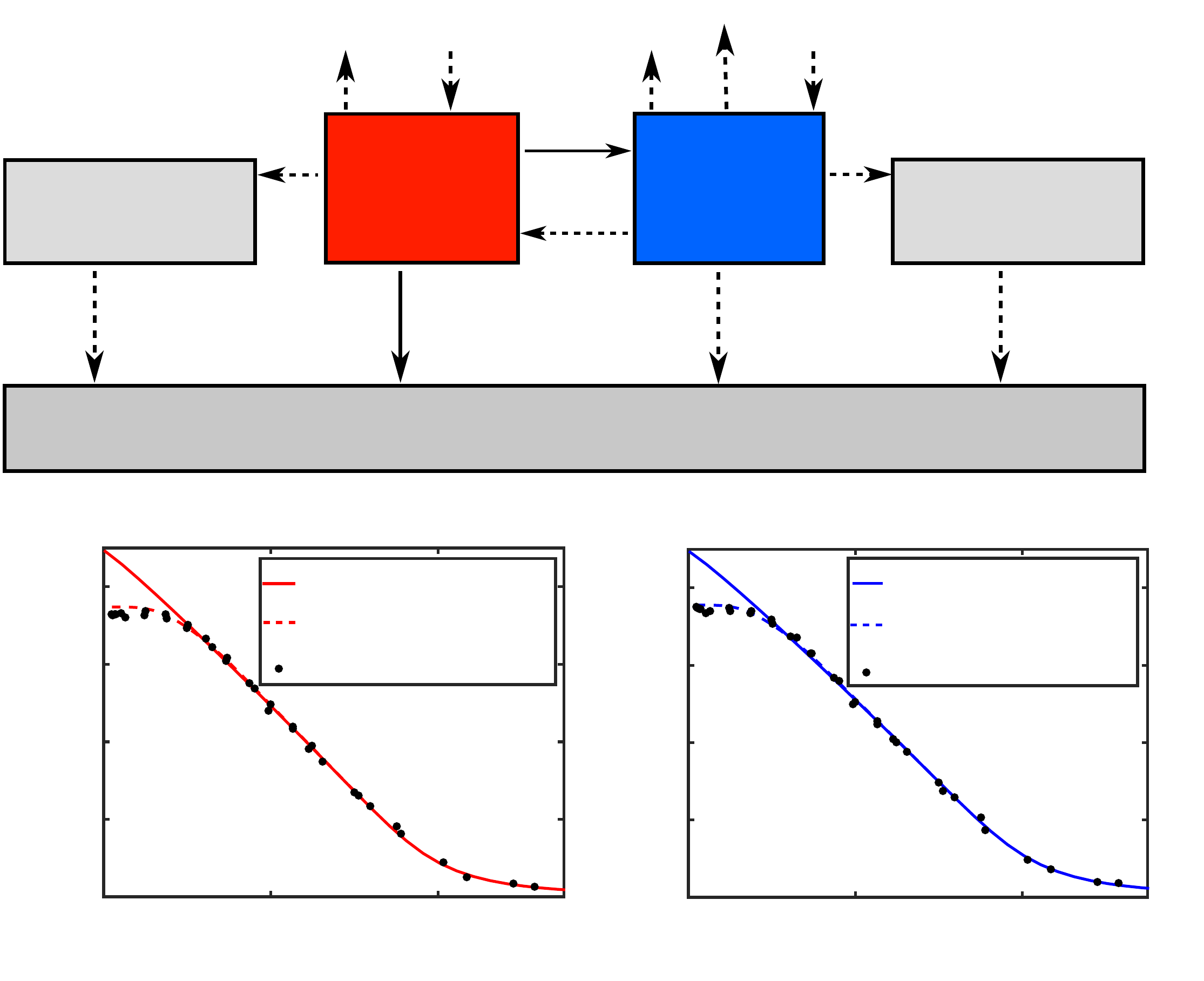
  }
  \caption{ \textbf{ Thermal model and temperature calibration.}
  \textbf{a},
  In the full thermal model, the conduction electron systems of  the normal-metal islands exchange heat with each other by photonic heat conduction, $P_\Gamma$, and  with the near-by normal-metal reservoirs through quasiparticle heat conduction, $P_{\textrm{qp},i}$. 
  The electrons exchange heat directly with  the phonon bath through the electron--phonon coupling $P_{\textrm{ep},i}$. 
  The temperature of Island~B is controlled with power of the NIS junction, $P_\textrm{NIS}$, and the thermometers introduce powers  $P_{\textrm{th},i}$.
  Heat leakages   from a high-temperature environment, $P_{\textrm{leak},i}$,  are also included in the model as well as a parasitic heat flow between the islands, $P_{\textrm{p}}$.
  Here, subscripts A, B, AR, and BR denote Islands A and B, and the corresponding reservoirs, respectively.
  The  simplified model accounts only for $P_\Gamma$ and $P_\textrm{ep,A}$, which are denoted by solid arrows. 
  See Methods for details.
  \textbf{b, c}, Thermometer voltages as defined in Figs.~\ref{fig:sample_structure}c, g for Island~A (\textbf{b}) and  B (\textbf{c}) as functions of the bath temperature in Sample~A1. 
  The solid lines show theoretical predictions without any free parameters assuming that the electron temperature is equal to the bath temperature.
  These curves are used for the voltage--temperature conversion. 
  The discrepancy between the data and  the model below 100 mK indicates the saturation of the electron temperature, confirmed by the dashed lines showing the theoretically predicted voltages with the electron temperature solved from the full thermal model. 
  The uncertainty in the measurements is of the order of the marker size.
  }
  \label{fig:thermal_model}
\end{figure}

All measurements are performed at millikelvin temperatures. 
The voltage-biased ($V_\textrm{B}$) NIS junctions at Island~B are used for temperature control whereas the current-biased ($I_\textrm{th,A}$, $I_\textrm{th,B}$) junctions at both islands are used for thermometry.
The thermometer calibration data for Sample A1 is  shown in   Fig.~\ref{fig:thermal_model}b, c.
Here, we  focus mainly on the analysis of Sample A1, which exhibits  the highest photonic thermal conductance according to the model.
It also achieves  the lowest electron temperature, below 90 mK at the 10-mK base temperature of the cryostat.
The other samples show higher minimum electron temperatures, which increases the uncertainty in their thermometry at the very low temperatures. 


We analyse the thermal conductance between the Islands A and B, $G_\textrm{AB}$, and those to the phonon bath, $G_\textrm{A0}$ and $G_\textrm{B0}$,  as schematically presented in Fig.~\ref{fig:sample_structure}h.
By linearising the heat flows of Island~A  at small temperature differences,  energy conservation yields a  differential temperature response:
$\textrm{d}T_\textrm{A}/\textrm{d}T_\textrm{B} = G_\textrm{AB}/( G_\textrm{AB}+  G_\textrm{A0})$. 
If the conductance $G_\textrm{AB}$ between the islands is quantum limited and the heat conduction to the bath stems from qualitatively different phenomena, $G_\textrm{AB}$ dominates over $G_\textrm{A0}$ at low enough temperatures.
Thus,  the temperature response generally tends to unity with decreasing temperatures.

The photonic net power flow from  normal-metal Island A to B is given by (Methods and Ref.~\onlinecite{Pascal_2011})
\begin{equation}
  P_\Gamma =\int_0^\infty \frac{\textrm{d} \omega}{2\pi} \hbar \omega |t(\omega)|^2 \left[\frac{1}{\exp(\frac{\hbar \omega}{k_\textrm{B} T_\textrm{A}}) - 1} - \frac{1}{\exp(\frac{\hbar \omega}{k_\textrm{B} T_\textrm{B}}) - 1} \right] \label{eq:Pphotonic}
\end{equation}
where $\hbar$ is the reduced Planck constant, $k_\textrm{B}$ is the Boltzmann constant, and $|t(\omega)|^2$ is a transmission coefficient that depends on the photon angular frequency $\omega$, the characteristic impedance of the transmission line, and the resistances of the terminating normal-metal islands (see Eq.~\ref{eq:transmissioncoeff} for details).
If the characteristic impedance of the transmission line  equals the island resistances, we have $|t(\omega)|^2 =1$. 
In this case, an analytical solution is obtained, $P_\Gamma = \frac{\pi k_\textrm{B}^2 }{12 \hbar} (T_\textrm{A}^2 - T_\textrm{B}^2)$,
which  can further  be expressed in terms of the quantum of thermal conductance as $P_\Gamma = G_\textrm{Q} (T_\textrm{A} - T_\textrm{B})$, where $G_\textrm{Q} = \frac{\pi k_\textrm{B}^2 T}{6 \hbar}$ with $T=(T_\textrm{A} + T_\textrm{B})/2$. 
The thermal conductance to the phonon bath at temperature $T_0$ can be approximated by the electron--phonon conductance as $G_\textrm{A0} \approx G_\textrm{ep,A} = 5 \Sigma_{\textrm{N}} \Omega_\textrm{A}  T_0^4$ in  Island~A (Methods).
Here,  $\Sigma_\textrm{N}$  is a material parameter describing the strength of the electron--phonon coupling in the normal metal, and $\Omega_\textrm{A}$ is the volume of Island~A.
Thus, one  obtains a simple theoretical prediction  without any free parameters based on  quantum-limited photonic heat conduction and electron--phonon coupling
\begin{equation}
 \frac{\textrm{d}T_\textrm{A}}{\textrm{d}T_\textrm{B}} = \frac{1}{1 + a T_0^3}
 \label{eq:simplified_theory}
\end{equation}
where  $a=30 \Sigma_\textrm{N} \Omega_\textrm{A} \hbar/(\pi k_\textrm{B}^2)$ is a predetermined constant.
We also devise a full thermal model shown in Fig.~\ref{fig:thermal_model}a for a more accurate description of  the  heat flows (Methods).



Figure~\ref{fig:full_simulation} shows the measured island temperatures for Sample~A1 at various bath temperatures together with the corresponding results from the full thermal model.
The experimental observations are in good agreement with the simulations over a wide range of bath temperatures and bias voltages.
The maximum cooling of Island~B is obtained at its bias voltages $V_\textrm{B} \lesssim 2\Delta/e \approx 0.45$~mV, where $\Delta$ is the superconductor energy gap and $e$ is the elementary charge.
The clearly observed cooling of distant Island A cannot be attributed to phonons since the phonon bath temperature increases as a result of the net heating power $| V_\textrm{B} I_\textrm{B} |$ dissipated during cooling of Island B.
In addition to photons, quasiparticles may contribute to the heat conduction, but their effect is weak and qualitatively in contradiction with our observations at low temperatures~\cite{Timofeev_1}. 
Furthermore, the essentially vanishing temperature response for the control sample in Fig.~\ref{fig:full_simulation}f indicates that the photonic channel dominates in the heat conduction between the islands in Sample~A1.

To accurately analyse the photonic heat conduction, we show the  temperature of Island~A in Fig.~\ref{fig:T2_simulation}a as a function of the electron temperature of Island~B for different bath temperatures.
The measured temperatures of Island A in Fig.~\ref{fig:T2_simulation}a are insensitive to heat conduction mechanisms that only involve Island~B and its reservoirs.
Therefore, this analysis method provides a robust way of studying photonic heat conduction.
In Sample A1, the curvatures of $T_\textrm{A}$ as a function of $T_\textrm{B}$ are negative, which is in stark contrast to the positive curvature observed for the  control sample.
This fundamental difference is due to the absense of the photonic heat conduction in the control sample.
In Samples A2 and A3 (Table~\ref{tab:samples}), the curvatures resemble that of A1 (data not shown).
At high bath temperatures, $T_\textrm{A}$ is almost independent of $T_\textrm{B}$, which is a consequence of the strong electron--phonon coupling.
Figure~\ref{fig:T2_simulation}b shows the differential temperature response, $\textrm{d}T_\textrm{A}/\textrm{d}T_\textrm{B}$, extracted at the lowest $T_\textrm{B}$  obtained for each bath temperature.
The steep increase in   $\textrm{d}T_\textrm{A}/\textrm{d}T_\textrm{B}$ at low bath temperatures  is a signature of the photonic heat conduction:
the thermal conductance between the islands, $G_\textrm{AB}$, determined by the photonic heat conduction, dominates over the conductance to the bath, $G_\textrm{A0}$.

\definecolor{magenta_1}{rgb}{1,0,1}

\begin{figure}[t!]
  \centering
  \def\svgwidth{89mm}
  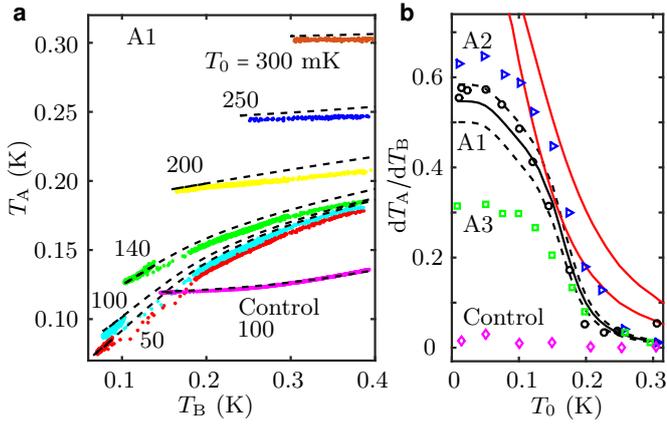
  \caption{ \textbf{ Differential temperature response and the quantum of thermal conductance. }
  \textbf{a}, Measured (dots) and simulated (dashed lines) temperatures of Island~A as functions of the temperature of Island~B for the indicated phonon bath temperatures in Sample A1. 
  The results for the control sample at 100 mK bath temperature are shown for comparison.
  \textbf{b}, Differential temperature response (\mbox{\boldmath${\pmb\circ}$}) from (\textbf{a}) at the lowest $T_\textrm{B}$ for each bath temperature. 
  The experimental uncertainty is of the order of the marker size.
  For comparison, we also show the corresponding experimental data for the control sample (\textcolor{magenta_1}{\mbox{\boldmath${\pmb \lozenge}$}}),  for Sample A2 (\textcolor{blue}{\boldmath${\pmb \triangleright}$}), and  for  A3 (\textcolor{green}{\boldmath${\pmb \medsquare}$}).
  The solid black line shows the prediction of the full thermal model of  Fig.~\ref{fig:thermal_model}a. 
  The dashed lines are calculated with 80~\% (bottom) and 115~\% (top) of the quantum of thermal conductance indicating the sensitivity of the results to the photonic heat conduction. 
  The solid  red lines are calculated with the simplified thermal model [Eq.~\eqref{eq:simplified_theory}] for Sample A1 without any free parameters: 
   we take into account only quantum-limited photonic heat conduction and  electron--phonon coupling with literature values \cite{Giazotto} for the   constant of Cu, 
   $\Sigma_\textrm{N} = 2\times10^9$~WK$^{-5}$m$^{-3}$ (right), and $\Sigma_\textrm{N} = 4\times10^9$~WK$^{-5}$m$^{-3}$ (left). 
  }
  \label{fig:T2_simulation}
\end{figure}



In addition,  Fig.~\ref{fig:T2_simulation}b shows a prediction of the simplified model according to Eq.~\ref{eq:simplified_theory}.
Despite its simplicity, it captures the essential features of the experimental data of Samples~A1 and~A2 which exhibit photonic heat conduction very close to the quantum limit, $G_\textrm{Q}$.
The deviation between the data and  the simplified model at high temperatures is due to the neglected  quasiparticle heat conduction between the islands and their reservoirs which increases $G_\textrm{A0}$.
At low temperatures, on the other hand, the discrepancy increases due to the saturation of the electron temperatures not present in the simplified model.
These observations bring insight to the good agreement between the experimental observations and the full thermal model.


In summary, we experimentally demonstrate  quantum-limited heat conduction over  macroscopic distances.
The on-chip design of the resistors enables their straightforward  utilisation in a multitude of different applications, including the initialization of quantum bits~\cite{Geerlings2013} by controlled cooling  and remote cooling of other quantum-technological components.
The methods developed in this study may also be used in the future to implement efficient heat transfer between separate chips and temperature stages of the cryostat. 
For example, the remotely cooled quantum device may be operated at a typical base temperature whereas the cold reservoir may be located at a lower-temperature stage which is incompatible with the relatively large power consumption of the actual device.
The  cooling  distance we demonstrate here   is sufficient for practically  all present-day applications.

\vspace{15 pt}
\begin{center}
\textbf{METHODS}
\end{center}


\vspace{1 pt}\noindent
\textbf{Sample fabrication}

\noindent
The samples are fabricated in a multi-step process on 0.5-mm-thick silicon wafers with 300-nm-thick thermally grown silicon oxide layers. 
The transmission lines are fabricated in an optical-lithography process using a mask aligner and an electron beam evaporator. 
The wafers are cleaned with reactive ion etching before the metal deposition. 
The Al film has a thickness of 200 nm, on top of which films of Ti and Au are deposited with thicknesses of 3 and 5 nm, respectively, to prevent oxidation.

The nanostructures are fabricated with electron beam lithography. 
The mask consists of poly(methyl methacrylate) and  poly[(methyl methacrylate)-\textit{co}-(methacrylic acid)] layers, which enable a large undercut necessary for three-angle shadow evaporation. 
Prior to the metal deposition, the samples are cleaned with argon plasma in the electron beam  evaporator.
As the first metal, we deposit an Al layer which is oxidised \textit{in situ}  introducing the insulator layer for the NIS junctions. 
Subsequently, a layer of normal metal is deposited  followed by a layer of Al. 
The normal metal is either AuPd (mass ratio 3:1) or Cu.
Lift-off of the excess metal is performed with acetone followed by cleaning with isopropanol.

\vspace{5 pt}\noindent
\textbf{Measurements}
\newline
\noindent
The electrical measurements are performed at millikelvin temperatures achieved with a commercial cryogen-free dilution refrigerator. 
The chip is attached to a sample holder containing a printed circuit board (PCB), to which the sample is connected by Al bond wires. 
The PCB is  connected to room-temperature measurement setup with lossy coaxial cables. 
To suppress electrical noise, the power-line-powered devices are connected to the sample through opto-isolators. 
Battery-powered amplifiers and voltage and current sources  are connected to the sample without  opto-isolation.
The voltage $V_\textrm{B}$ is swept slowly (down to 1 $\mu$V$/$s) to avoid apparent hysteresis. 
Furthermore, the measurements are repeated several times, and the data points with clear disturbance from random external fluctuations are  excluded.

\vspace{5 pt}
\noindent
\textbf{Photonic heat conduction } 
\newline
\noindent
Here, we derive  Eq.~\eqref{eq:Pphotonic} for the photonic heat conduction starting from the first-principles circuit quantum electrodynamics.
Previously, our case of two islands coupled with a transmission line has been studied with the help of classical circuit theory~\cite{Pascal_2011}.
These results can also be obtained using paht integrals~\cite{Golubev_2015}.
In contrast,  we analyse the system using methods discussed in ref.~\onlinecite{Yurke}.
In particular, the Heisenberg equations of motion are given by Kirchhoff's circuit laws.
Furthermore, a terminating resistor can be treated as a semi-infinite transmission line with a characteristic impedance equal to its resistance \cite{Yurke}.

\begin{figure}[t!]
  \centering
  \def\svgwidth{89mm}
  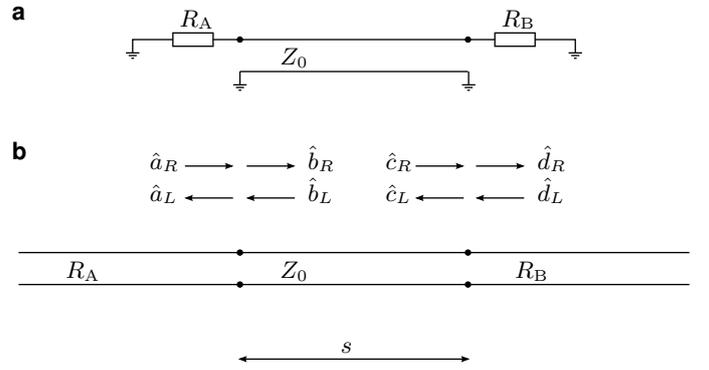
  \caption{ \textbf{Transmission line diagrams.}  
	    \textbf{a}, Circuit diagram of a transmission line with characteristic impedance $Z_0$  terminated by resistors $R_\textrm{A}$ and $R_\textrm{B}$. 
	    \textbf{b}, System in (\textbf{a}) represented  as   three connected transmission lines. 
	    The annihilation operators for the left- and right-moving photons at the different points of the system are denoted by $\hat a_i, \hat b_i,\hat c_i,$ and $\hat d_i,$ $i\in\{\textrm{L,R}\}$.
	    }
  \label{fig:Kuva_suppl_photonic}
\end{figure}

We  express the boundary conditions originating from the Kirchhoff's laws for the photon annihilation operators defined in  Fig.~\ref{fig:Kuva_suppl_photonic} as 
\begin{eqnarray}
 \frac{1}{\sqrt{Z_0}} ( \hat b_\textrm{L} -  \hat b_\textrm{R}) &=& -\frac{1}{\sqrt{R_\textrm{A}}} ( \hat a_\textrm{R} -  \hat a_\textrm{L}) \\
 \frac{1}{\sqrt{Z_0}} ( \hat c_\textrm{R} -  \hat c_\textrm{L}) &=& -\frac{1}{\sqrt{R_\textrm{B}}} ( \hat d_\textrm{L} -  \hat d_\textrm{R}) \\
 \sqrt{Z_0} ( \hat b_\textrm{L} +  \hat b_\textrm{R}) &=&  \sqrt{R_\textrm{A}} (   \hat a_\textrm{R} +    \hat a_\textrm{L}) \\
 \sqrt{Z_0} ( \hat c_\textrm{R} +  \hat c_\textrm{L}) &=&  \sqrt{R_\textrm{B}} (   \hat d_\textrm{L} +    \hat d_\textrm{R}) \\
  \hat c_\textrm{R} &=& e^{i \phi}  \hat b_\textrm{R} \\
  \hat b_\textrm{L} &=& e^{i \phi}  \hat c_\textrm{L}
\end{eqnarray}
where $\phi = \omega s/v$ is the phase shift obtained by a wave with angular frequency $\omega$ and velocity $v$ when travelling over distance $s$.
Assuming no photons coming from the right, $\hat d_\textrm{L}=0$, we can solve the transmission coefficient $t$ defined as $\hat d_\textrm{R} = t(\omega) \hat a_\textrm{R}$.
Thus, we obtain
\begin{equation}
 |t(\omega)|^2=\frac{2}{1+\frac{R_\textrm{A}^2+R_\textrm{B}^2}{2 R_\textrm{A} R_\textrm{B}}  + \frac{R_\textrm{A}^2R_\textrm{B}^2 +Z_0^4- R_\textrm{A}^2 Z_0^2- R_\textrm{B}^2 Z_0^2}{2 R_\textrm{A} R_\textrm{B} Z_0^2}  \sin^2(\phi)  }
 \label{eq:transmissioncoeff} 
\end{equation}
The transmission coefficient is symmetric with respect to the exchange of resistances $R_\textrm{A}$ and $R_\textrm{B}$.
In a matched case, $R_\textrm{A}=R_\textrm{B}=Z_0$, Eq.~\eqref{eq:transmissioncoeff} simplifies to $|t(\omega)|^2=1$.

Energy dissipation at the resistor $R_\textrm{B}$ can be obtained from the average  photon flux to the right in the transmission line with a characteristic impedance $R_\textrm{B}$ multiplied by the energy carried by each photon.
Here, the zero-point energy does not appear in the dissipated power.
Thus, the power per unit frequency can be expressed as
\begin{eqnarray}
 P_{\rightarrow,\omega} &=& \langle \hbar \omega \hat d_\textrm{R}^\dagger \hat d_\textrm{R}  \rangle \nonumber  \\
  &=& \hbar \omega |t(\omega)|^2 \langle  \hat a_\textrm{R}^\dagger \hat a_\textrm{R}  \rangle \nonumber   \\
  &=& \hbar \omega |t(\omega)|^2 \frac{1}{\exp(\frac{\hbar \omega}{k_B T_\textrm{A}}) - 1}
\end{eqnarray}
since the number of photons emitted by the left resistor is given in thermal equilibrium by the Bose--Einstein distribution.
Due to  symmetry, the power transfer to the opposite direction is given by
\begin{eqnarray}
 P_{\leftarrow,\omega} &=& \hbar \omega |t(\omega)|^2 \frac{1}{\exp(\frac{\hbar \omega}{k_B T_\textrm{B}}) - 1}
\end{eqnarray}
The net photonic heat transport from  $R_\textrm{A}$ to $R_\textrm{B}$ is, therefore, given by 
\begin{align}
 P_\Gamma \!\!  &= \!\! \!\!  \int_0^\infty \frac{\textrm{d}\omega}{2\pi}( P_{\rightarrow,\omega} - P_{\leftarrow,\omega} ) \nonumber \\
 &= \!\! \!\! \int_0^\infty \frac{\textrm{d}\omega}{2\pi} \hbar \omega |t(\omega)|^2 \left[\frac{1}{\exp(\frac{\hbar \omega}{k_B T_\textrm{A}}) - 1} - \frac{1}{\exp(\frac{\hbar \omega}{k_B T_\textrm{B}}) - 1} \right] 
\end{align}
In the special case  of a vanishing waveguide length, $s\rightarrow 0$,   Eq.~\eqref{eq:transmissioncoeff} yields $|t(\omega)|^2 = 4 R_\textrm{A} R_\textrm{B} / (R_\textrm{A}+R_\textrm{B})^2$ 
which is identical to the result considered in ref.~\onlinecite{Schmidt2004}  for two resistors in a loop.
On the other hand, if one sets $Z_0$ to be inversely proportional to $s$ and takes the limit $s\rightarrow 0$, 
one obtains $|t(\omega)|^2 = 4 R_\textrm{A} R_\textrm{B} / [(R_\textrm{A}+R_\textrm{B})^2 + X^2 ]$ with a reactance $X=Z_0 \omega s/v$.
This result reproduces that of two resistances connected in a loop with a series reactance~\cite{Schmidt2004, Pascal_2011}.

\vspace{10 pt}\noindent
\textbf{NIS thermometry}

\noindent
The quasiparticle  current through an NIS junction with tunnelling resistance $R_\textrm{T}$ is given in the sequential-tunnelling theory by \cite{Giazotto} 
\begin{equation}
I(V,T_\textrm{N})  \!\!  = \!\!   \frac{1}{e R_\textrm{T} }  \!\!  \int^{\infty}_{0}  \!\!  n_{S}(E) [f(E-eV, T_\textrm{N})-f(E+eV, T_\textrm{N})] \textrm{d}E
\label{eq:NIS_current}
\end{equation}
where $T_\textrm{N}$ is the normal-metal electron temperature, and $V$ the voltage across the junction.
Here, the Fermi--Dirac distribution is given by
\begin{equation}
 f(E, T) = \frac{1}{e^{E/(k_\textrm{B} T)}+1}
 \label{eq:fermi-dirac}
\end{equation}
and the superconductor density of quasiparticle states  assumes the form
\begin{equation}
n_\textrm{S}(E) = \left| \textrm{Re} \frac{E/\Delta+i\gamma}{\sqrt{ (E/\Delta +i\gamma)^2-1}} \right|
\label{eq:Dynes}
\end{equation}
Above, $\gamma$ is the Dynes parameter~\cite{Giazotto} accounting for  the subgap current, and $\Delta$ is the superconductor energy gap. 
Experimentally, $\gamma$ is obtained as the ratio of the asymptotic resistance at large voltages and the resistance at zero voltage.
We note that Eq.~\eqref{eq:NIS_current} has  a very weak dependence on the temperature of the superconductor through the temperature dependence of $\Delta$.
Thus, an NIS junction can be used as a thermometer probing the electron temperature of the normal-metal.
We apply a constant current, and deduce the temperature from the measured voltage according to a calibration curve shown in Fig.~\ref{fig:thermal_model}.

\vspace{10 pt}
\noindent
\textbf{Thermal model}

\noindent
In the full thermal model illustrated in Fig.~\ref{fig:thermal_model}a, we consider several heat transfer mechanisms:
Firstly, the NIS junctions produce heat flows between the normal-metal islands and the superconducting leads.
Secondly, the electrons in the normal metal exchange heat with the phonon bath. 
Thirdly, the  islands exchange heat with each other by photons travelling in the transmission line.
Finally, the model takes into account geometrical properties of the samples as well as properties specific to the measurement setup.

The NIS junctions can be used for cooling and heating of the normal metal, and the power out of the normal metal can be  computed from \cite{Giazotto}
\begin{equation}
P_\textrm{ideal}  \!\!  =   \!\! \frac{1}{e^2 R_\textrm{T} }   \int_{-\infty}^{\infty}     \!\!   \!  n_\textrm{S}(E) (E \! - \! eV)[f(E \! - \! eV, T_\textrm{N}) \! - \! f(E, T_\textrm{S})]\textrm{d}E
\label{eq:Prefr_NIS_1}
\end{equation}
We model the nonidealities in the NIS  power by assuming a constant fraction, $\beta$, of the power flowing to the superconductor to flow back to the normal metal. 
Thus, the back-flow power can be written as
\begin{equation}
 P_\textrm{bf}=\beta(IV + P_\textrm{ideal})
 \label{eq:P_bf_beta}
\end{equation}
where $IV$ gives the total power.
Consequently, the total cooling power of an NIS junction is given by
\begin{equation}
 P_\textrm{NIS}=P_\textrm{ideal} - P_\textrm{bf}
 \label{eq:P_NIS}
\end{equation}
The physical background for the back flow has been studied in ref.~\onlinecite{ONeil_beta}.
A factor of 2 is included in the power when two NIS junctions are connected to form an SINIS structure. 
Since the thermometers are based on similar NIS junctions, their powers are calculated with the same equations as for the actual  power used to control the temperature of Island~B. 
However, the voltages across the thermometer junctions must first be solved using Eq.~\eqref{eq:NIS_current}, the island temperature and the thermometer bias current.

The electrons in the normal metal are coupled to the phonon bath, and the heat flow is given by \cite{Giazotto}
\begin{equation}
P_{\textrm{ep},i}= \Sigma_{\textrm{N}} \Omega_i ( T_i^5 - T_0^5 )
\label{eq:Pep_theor}
\end{equation}
Here,   $\Omega_i$ is the volume of  normal-metal block $i \in \{\textrm{A, B, AR, BR}\}$. 
For Cu and AuPd, the parameter $\Sigma_\textrm{N}$ is typically~\cite{Timofeev_1, Giazotto} between $2\times 10^{9}$ and $4\times 10^{9}$~WK$^{-5}$m$^{-3}$.
In the simulations, we use values $2.0\times 10^{9}$~WK$^{-5}$m$^{-3}$ and $3.0\times 10^{9}$~WK$^{-5}$m$^{-3}$ for Cu and AuPd, respectively, unless otherwise mentioned.
The normal metal under the  superconductors at the ends of the islands are  excluded from the volume in the simulations due to the superconductor proximity effect.
For small temperature differences, $T_i \approx T_0$, one obtains 
\begin{equation}
 P_{\textrm{ep},i}= G_{\textrm{ep},i} ( T_i - T_0 )
\end{equation}
where $G_{\textrm{ep},i} = 5 \Sigma_{\textrm{N}} \Omega_i  T_0^4$.

We account for heat leaks from a high-temperature environment by including constant heating powers to both islands, $P_\textrm{leak,A}$ and $P_\textrm{leak,B}$. 
They are fixed by the saturation of the electron temperature observed in  Fig.~\ref{fig:thermal_model} at low bath temperatures.

In the thermal model, we consider quasiparticle heat conduction only from the islands to  their near-by normal-metal reservoirs.
The reservoirs are a consequence of the three-angle evaporation method, and they provide an additional channel for thermalisation to the phonon bath.
At both islands, there are actually two reservoirs which are presented as one in  Fig.~\ref{fig:thermal_model}a for simplicity.
The extremely weak quasiparticle heat conduction from one island to the other over a distance longer than 5~mm is included in the parasitic heat conduction as discussed below.
The power flow at the normal-metal block $i$  due to the quasiparticles is given by
\cite{Timofeev_1}
\begin{equation}
P_{\textrm{qp},i} = \kappa_\textrm{S} A T'(x_i)
\label{eq:QP1}
\end{equation}
where $T'(x_i)$ is the derivative of the temperature in the superconductor with respect to the position coordinate $x_i$, and $A$ is the cross section of the line. 
The superconductor heat conductivity, $\kappa_\textrm{S}$, is related to the normal-state heat conductivity, $\kappa_\textrm{N}$, at a temperature $T$ by 
 \cite{Bardeen_conductivity}  
\begin{equation}
\kappa_\textrm{S}= \tilde{\gamma}(T) \kappa_\textrm{N}
\label{eq:kappa-s}
\end{equation}
where $\tilde{\gamma}$ is a  suppression factor 
\begin{equation}
\tilde{\gamma}(T) = \frac{3}{2 \pi^2} \int_{\Delta / (k_\textrm{B} T)}^{\infty} \frac{t^2}{\cosh^2(t/2)}\textrm{d}t
\label{eq:gamma_suppression}
\end{equation}
The normal-state heat conductivity of the line is obtained from the Wiedemann--Franz law as 
\begin{equation}
\kappa_\textrm{N} =  \frac{ L_0 T(x)}{\rho }
\label{eq:kappa-n}
\end{equation}
where $\rho$ is the normal-state electric resistivity of the line,   and $L_0=2.4\times 10^{-8}$~W$\Omega$K$^2$ is the Lorenz number.
The temperature profile in the superconducting lines can be calculated using a heat diffusion equation \cite{Timofeev_1}.
However, the electron--phonon coupling in a superconducting state is greatly suppressed with respect to that of a normal-state \cite{Timofeev_2}.
Thus, we neglect the electron--phonon coupling in the leads and assume here a linear temperature profile.

Andreev current plays a minor role in our experiments since the induced temperature changes at the islands are small in the subgap voltage regime where it may dominate \cite{Rajauria_Andreevcurrent}.
Therefore, we do not consider it in the thermal model.

We observe a weak island-to-island heat transport also in the control sample, in which the centre conductor is shunted as shown in Fig.~\ref{fig:Control_sample_SEM}.
We model this parasitic heat transport by letting a constant proportion, $\alpha$, of the total input power  at Island~B to flow into Island~A,
\begin{equation}
 P_\textrm{p} = \alpha I_\textrm{B} V_\textrm{B}
\end{equation}
The exact mechanism of the parasitic channel remains unknown, and the heat flow may depend on the sample geometry. 
The  parasitic heat conduction  extracted from the control sample includes all the heat conduction channels from one island to the other except the photonic heat conduction which is essentially absent due to the shunt.
This heat flow may be attributed to quasiparticles since they can travel long distances before recombination, especially at low bath temperatures.
Furthermore, although the electric contact of the shunting metal block between the ground plane and the centre conductor is of very low impedance,  small residual photonic heat conduction cannot be fully excluded.
Nevertheless, the parasitic heat conduction is much weaker than the total heat conduction in the actual devices.
We note that the parasitic heat conduction  is only added to the model for more accurate description at high heating powers.

We solve the heat balance equations for both islands and both reservoirs simultaneously. 
The equations can be expressed as (Fig.~\ref{fig:thermal_model}a)
\begin{align}
		   P_\Gamma  + P_\textrm{th,A} -  P_\textrm{leak,A} -  P_\textrm{p} +  P_\textrm{ep,A}  +  P_\textrm{qp,A} &= 0 \\  
  \!\!  P_\textrm{NIS} -  P_\Gamma  + P_\textrm{th,B} -  P_\textrm{leak,B} +  P_\textrm{p} +  P_\textrm{ep,B}  +  P_\textrm{qp,B} &= 0 \\
										       P_\textrm{ep,AR} -  P_\textrm{qp,A} &= 0 \\
										       P_\textrm{ep,BR} -  P_\textrm{qp,B} &= 0 
\end{align}
These equations yield the temperatures $T_i$,  $i \in \{\textrm{A, B, AR, BR}\}$  for a given phonon bath temperature, $T_0$, and bias voltage, $V_\textrm{B}$, both of which are accurately controlled.

\begin{table*}[t!]
 \begin{center}
 \caption{ \textbf{Simulation parameters for Sample A1.} 
 Sample dimesions are based on the actual sample, 
 $I_\textrm{th,A}$ and $I_\textrm{th,B}	$ are externally controlled, 
 $\Delta$, $\gamma$ and  resistances are extracted from independent current--voltage characteristics, 
 $\rho$ assumes a typical value for evaporated aluminium,
 $L_l$ is calculated  analytically \cite{Goppl},
 $C_l$ is based on a finite-element simulation, 
 $P_\textrm{leak,A}$ and  $P_\textrm{leak,B}$ are obtained from island temperature saturation at low phonon bath temperatures in  Fig.~\ref{fig:thermal_model},
 $\alpha$ is extraced from the control sample,
 $\beta$ is obtained from the nonideal cooling power of the NIS junctions,
 $\Omega_\textrm{AR}$ and $\Omega_\textrm{BR}$ are the actual  volumes multiplied by 2 to account also for the quasiparticle recombination in the superconductors,
 and $T_\textrm{satur}$ and $T_\textrm{const}$ are fitting parameters assuming realistic values.
 }
 \begin{tabular}{ l | c |c | c  }
   \hline
Parameter	&		Symbol		&		Value		&	Unit				\\
\hline														
\hline														
														
	 Island volume	&	$	\Omega_\textrm{A},$ $ \Omega_\textrm{B}	$	&	$	4500 \times 300 \times 20       	$	&	nm$^3$				\\
	 Effective reservoir  volume	&	$	\Omega_\textrm{AR}, \Omega_\textrm{BR}	$	&	$	10 \times  \Omega_\textrm{A} 	$	&					\\
	 Cross section of   lines from island to reservoir	&	$	A	$	&	$	300 \times 100    	$	&	nm$^2$				\\
	Distance from island to reservoir	&	$	l_\textrm{IR}	$	&	$	24    	$	&	$\mu$m				\\
	 Waveguide length	&	$	s	$	&	$	0.193             	$	&	m				\\
	Superconductor energy gap	&	$	\Delta	$	&	$	224     	$	&	$\mu$eV				\\
	 Inductance per unit length	&	$	L_l	$	&	$	4.14 \times 10^{-7}	$	&	H$/$m				\\
	Capacitance per unit length	&	$	C_l	$	&	$	1.51 \times 10^{-10}     	$	&	F$/$m				\\
	Thermometer bias current	&	$	I_\textrm{th,A},$ $I_\textrm{th,B}	$	&	$	18   	$	&	pA				\\
	 Resistivity of Al lines in normal state	&	$	\rho 	$	&	$	 1.0 \times 10^{-8}          	$	&	$\Omega$m				\\
	 Dynes parameter	&	$	\gamma	$	&	$	1.05 \times 10^{-4}      	$	&					\\
	 Material parameter for Cu	&	$	\Sigma_\textrm{N}	$	&	$	2.0  \times 10^{9}           	$	&	W K$^{-5}$ m$^{-3}$				\\
	 Normal state junction resistance	&	$	R_\textrm{T}	$	&	$	15.5           	$	&	k$\Omega$				\\
	 Island  resistance	&	$	R_\textrm{A},$ $R_\textrm{B}	$	&	$	65              	$	&	$\Omega$				\\
	 Characteristic impedance 	&	$	Z_0	$	&	$	\sqrt{ L_l/C_l}	$	&					\\
	 Phase velocity	&	$	v	$	&	$	1/\sqrt{ C_l L_l}	$	&					\\
	 Lorenz number	&	$	L_0	$	&	$	2.4  \times 10^{-8}          	$	&	W $\Omega$ K$^{-2}$				\\
	Saturation quasiparticle temperature for heat conductivity	&	$	T_\textrm{satur}	$	&	$	0.184	$	&	K				\\
	Additional quasiparticle temperature for heat conductivity	&	$	T_\textrm{const}	$	&	$	0.036	$	&	K				\\
	Parasitic heat conduction parameter	&	$	\alpha	$	&	$	3 \times 10^{-4}	$	&					\\
	 NIS back flow constant	&	$	\beta	$	&	$	0.056	$	&					\\
	Heat leak	&	$	P_\textrm{leak,A},$  $P_\textrm{leak,B}	$	&	$	1.6	$	&	fW				\\

\hline
\end{tabular}
 \label{tab:simulation_parameters}
 \end{center}
\end{table*}

The  parameters used in the full thermal model are shown in  Table \ref{tab:simulation_parameters}.
In the simulations, we slightly adjust the quasiparticle heat conductivity for improved agreement between the model and the experiments. 
More specifically, we set the temperature in Eq.~\eqref{eq:gamma_suppression} to be equal to the island temperatures increased by a small constant value and, in addition, we set the suppression factor to saturate at low temperatures. 
Hence, we introduce a replacement $\tilde\gamma(T) \rightarrow \tilde\gamma(T+T_\textrm{const}) + \tilde\gamma(T_\textrm{satur})$.
This approximation can be justified by several arguments.
Firstly, the superconductor heat conductivity depends on the purity of the sample \cite{Satterthwaite_Aluminium}.
Secondly, the superconductor energy gap has been observed to increase at small film thicknesses \cite{Court_Al_gap}. 
We use for all superconductors the same value, which is obtained from the current--voltage measurements of the NIS junctions,  although the leads are thicker.
The possibly smaller actual energy gap effectively corresponds to higher temperatures.
Thirdly, the neglected electron--phonon coupling in the superconducting leads may result in nonlinear temperature profile  increasing the quasiparticle heat conduction. 
The impurities in the sample may increase the electron--phonon coupling.
Fourthly, the heat leakage through the measurement cables from a high-temperature environment and other possible heat leak mechanisms may increase the temperature of the superconductors. 
Increased quasiparticle densities have been observed previously, and they can be suppressed by effective shielding and enhanced relaxation \cite{Saira_quasiparticle}.
Weak quasiparticle recombination can induce elevated quasiparticle temperatures.
However, we  increase only the heat conductivity and consider a  linear temperature profile in the lead between the island and the near-by reservoir.
In the simulations, the reservoirs have  effective volumes somewhat larger than their physical volumes, thus, taking into account the quasiparticles thermalising in the reservoirs and the ones recombining in the  superconductors. 
The requirement of the effective volume may also  be explained by the uncertainty in the employed literature value of the  electron--phonon coupling constant~\cite{Giazotto}.

\vspace{10 pt}\noindent
\textbf{Additional control samples without resistors}

\noindent 
We also fabricated and measured control samples without the normal-metal resistors terminating the transmission line. 
Instead, the transmission line is connected to input and output ports through coupling capacitors. 
The capacitors and the transmission line form a resonator, the quality factor of which can be measured. Using an LCR model \cite{Goppl}, we  extract the internal quality factor of the system to be of the order of 
60,000
indicating a negligibly weak effect in the heat conduction experiments. 
In fact, the energy losses due to the observed finite quality factor would only limit the photonic heat conduction beyond distances of the order of a kilometer.


\bibliographystyle{naturemag}
\bibliography{bibliography_1}



\textbf{Acknowledgements} 
We acknowledge the provision of facilities and technical support by Aalto University at Micronova Nanofabrication Centre.
We also  acknowledge funding  by the European Research Council under Starting Independent Researcher Grant No. 278117 (SINGLEOUT), the Academy of Finland through its Centres of Excellence Program (grant no.\ 251748) and grants (nos\ 138903, 135794, 265675, 272806 and 276528),  the Emil Aaltonen Foundation,  the Jenny and Antti Wihuri Foundation and the Finnish Cultural Foundation. 
We thank M. Meschke, J. P. Pekola, D. S. Golubev, M. Kaivola and J. C. Cuevas for discussions concerning this work. 
Furthermore, we thank L. Gr\"onberg for assistance in sample fabrication, and E. Mykk\"anen and A. Kemppinen for assistance in measurements.

\textbf{Author Contributions} M.P. and K.Y.T. fabricated the samples, developed and conducted the experiments, and analysed the data. 
J.G., R.E.L. and M.K.M. assisted in the sample fabrication and measurements.
T.T. contributed to the measurements.
M.M. provided the initial ideas and suggestions for the experiment and supervised the work in all respects. 
All authors discussed both experimental and theoretical results and commented on the manuscript that was written by M.P. and M.M.

\textbf{Author Information} 
Correspondence and requests for materials should be addressed to M.M.\ (mikko.mottonen@aalto.fi).

\nuke{\begin{center}
\textbf{METHODS}
\end{center}

\note{Full methods go here.}}

\clearpage
%
%
%
%


\clearpage


\clearpage

%
%
%
%
%
%
%
%
%
%

\end{document}